\documentstyle[12pt]{article}
\input epsf
\title{\bf Bose-Einstein condensation of a quantum group boson gas}
\author {\\ \\ \\ Marcelo R. Ubriaco\thanks{E-mail:ubriaco@ltp.upr.clu.edu}\\
{\em Laboratory of Theoretical Physics}\\
{\em Department of Physics}\\
{\em  University of Puerto Rico}\\
{\em P. O. Box 23343, R\'{\i}o Piedras}\\
{\em PR 00931-3343, USA}}
\date{}
\begin{document} 
\vspace{0.3in}
\maketitle
\vspace{0.15in}
\begin{abstract}
We study the Bose-Einstein condensation of a gas with $SU_q(2)$ symmetry. 
We show, in the thermodynamic limit, that the boson interactions introduced by the
quantum group symmetries enhance Bose-Einstein condensation giving a
discontinuity in the heat capacity $C_v$ at the critical temperature
$T_c$. The critical
temperature and the gap in $C_v$ increase with the value of the
parameter $q$ and become approximately constant for $q>3$.
 
\end{abstract}
\vspace*{.2in}
PACS number(s): 05.30.-d
\newpage
\baselineskip20pt
\section{Introduction}

The search for new applications 
 of quantum groups \cite{Jimbo,Ch}, other than the theory of integrable models, have diversified to
several areas of theoretical physics.  The literature
on this subject  deals mainly with formulations to build quantum group 
versions of the
Lorentz and Poincar\'{e} algebras \cite{CW}, or the use of a
quantum group as an internal symmetry 
in quantum mechanics and field theory \cite{AV}.
Although many of these approaches address interesting theoretical
questions, they in general remain at a formalism level, with the quantum group
parameter $q$ playing merely the role of a deformation parameter.
As it is well known, in the quantum inverse scattering method 
and vertex models, the parameter $q$ acquires a physical
meaning  through its relation with Planck's constant and
the anisotropy of the lattice respectively. 
In order to look for the physical role that $q$ could play in other areas of
physics,
in  previous papers \cite{U1,U2} we began a study of the thermodynamic properties
of quantum group gases, which are the quantum group fermion (QGF) and
quantum group boson (QGB) models. These models can be interpreted
as either fermion or boson gases with interactions fixed by the quantum group. 
For reasons of simplicity, we considered for both cases the simplest $SU_q(2)$ 
invariant Hamiltonian. In particular, the QGB model
has the interesting property that in two and three spatial dimensions
the parameter $q$ interpolates within a wide range of attractive and
repulsive systems including the free boson and fermion cases.  Therefore,
the departure from the value $q=1$, which at the mathematical level implies
the noncommutativity of quantum group matrix elements, gives
in a simple thermodynamic model, an alternative approach to fractional statistics.  

This work is a study on
the implications of introducing quantum group symmetries in a thermodynamic
system at low temperatures, with particular emphasis
in the phenomenon of Bose-Einstein condensation. There is, at present,
no direct indication that quantum group symmetries are realized in
any particular thermodynamic system.  At the moment, 
a quantum group gas is
 a  mathematical model which may or may not explain the behavior
of real gases in a particular situation.  Nevertheless, 
 by performing the kind of
calculations considered here, questions regarding the possible
relevance 
of quantum groups in a thermodynamic system can be answered.

Our starting point is to consider the simplest quantum group
, $SU_q(2)$, invariant Hamiltonian, and calculate the thermodynamic
properties of the corresponding interacting boson gas.  Our main motivation
resides in finding out about the role that these interactions, which result from the
requirement of quantum group invariance, may play at low temperatures.  It would be
of much interest, in the author's opinion, if these interactions could fit energy related
data obtained from the experiments.  

In experimental settings, 
trapping potentials are well approximated by the
potential of a harmonic oscillator. A noninteracting Bose gas in a harmonic potential \cite{GHT} is
known to exhibit a discontinuity in the heat capacity.  This discontinuity is, however,
considerably larger than the one reported in a recent experiment involving
a dilute gas of $^{87}Rb$ \cite{E}.
In addition, introducing  two-particle interactions, has the effect \cite{BPK} 
, for the case of a harmonic oscillator potential, of making the
heat capacity continuous, and therefore less agreeable with Reference \cite{E}. Therefore, it is natural 
to consider whether 
the boson interactions, required by the quantum group invariance of
the Hamiltonian, give a thermodynamic behavior that could fit this and other experimental data.  
 In addition, we expect 
these interactions to be weak, and a good agreement should occur for values of
$q$ close to the standard value $q=1$.  
 
In this paper, as a first step, we  study the QGB model
with no external potential.  An obvious question to address is, under which conditions
the QGB model exhibits Bose-Einstein condensation (BEC).  Under these conditions,
what is the behavior of the internal energy and heat capacity 
at the critical temperature $T_c$, and what is the dependence, if any, of $T_c$ 
on $q$. In this paper we answer those questions by analyzing 
the low temperature behavior of the QGB model
with $SU_q(2)$ symmetry.

In Section (\ref{ba}) we discuss the general
formalism and redefine our model  in terms of boson operators.  A more detailed 
and general discussion
is given in Reference \cite{U2}.  Section \ref{LT}  contains the main results
 of the paper.  We calculate the dependence of the heat capacity $C_v$, internal
energy $U$, entropy $S$ and equation of state on the parameter $q$.  
In particular, we show,
that the boson interactions  that arise from
the $SU_q(2)$ symmetry of the Hamiltonian, produce a
discontinuity, $\lambda$-transition, in the heat capacity at the
condensation temperature.
\section{$SU_q(2)$ bosons} \label{ba}

The quantum group $SU_q(2)$ consists of the set of matrices 
$T=\left(\begin{array}{cc} a & b \\ c & d\end{array}\right)$
with elements $\{a,b,c,d\}$ generating the algebra
\begin{eqnarray}
ab=q^{-1}ba  & , & ac=q^{-1}ca \nonumber \\
bc=cb & , & dc=qcd  \nonumber \\
db=qbd & , &  da-ad=(q-q^{-1})bc  \nonumber \\
& & det_{q}T\equiv ad-q^{-1}bc=1 ,
\end{eqnarray} 
with the unitary conditions \cite{VWZ} $\overline{a}=d, \overline{b}=q^{-1}c$
and $q\in {\bf R}$. Hereafter, we take $0\leq q<\infty$.
The $SU_q(2)$ transformation $T$ matrix and the corresponding $R$ matrix 

$$R=\left(\begin{array}{cccc} q & 0 & 0 & 0 \\ 0 & 1 & 0 & 0 \\
0 & q-q^{-1} & 1 & 0 \\ 0 & 0 & 0 & q\end{array}\right)$$

satisfy the algebraic relations \cite{Ta}
\begin{equation}
RT_1T_2=T_2T_1R,\label{T},
\end{equation}
and
\begin{equation}
R_{12}R_{13}R_{23}=R_{23}R_{13}R_{12},
\end{equation}
where $T_1=T\otimes 1$ and $T_2=1\otimes T$
$\in V\otimes V$, $(R_{23})_{ijk,i'j'k'}=
\delta_{ii'} R_{jk,j'k'} \in V\otimes V\otimes V$ .

We define as $SU_q(2)$-bosons the set of fields $\Phi=\left(\begin{array}{c} \Phi_1 \\ \Phi_2\end{array}\right)$,
satisfying the   relations
\begin{eqnarray}
\Phi_2\overline{\Phi}_2-q^2\overline{\Phi}_2\Phi_2&=&1 \label{1}\\
\Phi_1\overline{\Phi}_1-q^2\overline{\Phi}_1\Phi_1&=&1+(q^2-1)\overline{\Phi}_2\Phi_2
\\
\Phi_2\Phi_1&=&q\Phi_1\Phi_2\\
\Phi_2\overline{\Phi}_1&=&q\overline{\Phi}_1\Phi_2 \label{4}.
\end{eqnarray}

 Equations (\ref{1})-(\ref{4}) are covariant under the field redefinitions
 $\Phi'=T\Phi$ and $\overline{\Phi}'=\overline{\Phi}\:\overline{T}$ with $T\in SU_q(2)$.  
It is clear that for $q=1$, the operators
$\Phi_j$ become ordinary bosons $\phi_j$.

The operators $\Phi_j$ should not be confused with the so called $q$-boson oscillators. 
A set $(a_i,a_i^{\dagger})$ of $q$-bosons are defined, by the
relations \cite{Mf,B}
\begin{equation}
a_i a_i^{\dagger}-q^{-1}a_i^{\dagger}a_i=q^N ,\;\;\;
[a_i,a_j^{\dagger}]=0=[a_i,a_j],\label{q1}
\end{equation}
where $N|n\rangle=n|n\rangle$.  Several variations of Equation (\ref{q1})
are common in the literature.  By taking two sets of $q$-bosons, as defined in
Equation (\ref{q1}), it has been shown \cite{Mf,Ng}
that the operators 
\begin{equation}
J_+=a_2^{\dagger}a_1\;,\;\;J_-=a_1^{\dagger}a_2\;,\;\;
2J_3=N_2-N_1,
\end{equation}
provide a realization of the quantum Lie algebra $su_q(2)$
\begin{equation}
[J_3,J_{\pm}]=\pm J_{\pm}\;,\;\;\;\;
[J_-,J_-]=[2J_3].
\end{equation}

The main distinction between the $\Phi_j$ operators with $q$-bosons 
is that, in
contrast to Equations (\ref{1})-(\ref{4}),
the algebraic relations in Equation (\ref{q1}) are not covariant under the action
of the $SU_q(N)$ quantum group matrices.  Several studies on the thermodynamics of $q$-bosons, and similar
operators called quons \cite{G}, have been published \cite{TH,An}. In particular, 
a system defined by a 'free' Hamiltonian in terms of $q$-oscillators  
has been shown to exhibit BEC \cite{Mo,Tu}.   Certainly,
the work devoted to the thermodynamics of $q$-oscillators  represents a test
on the consequences of modifying  boson commutators, according to Equation (\ref{q1}) and
its different versions.  On the other hand, the algebraic relations in
Equations (\ref{1})-(\ref{4}), with the model discussed in this paper, 
address the implications that result of imposing
quantum group invariance in a thermodynamic system.

The simplest Hamiltonian written in terms of the operators $\Phi_j$ is certainly the one that
 becomes for $q=1$ the free boson
Hamiltonian with two species.  It is simply written as
\begin{equation}
{\cal H}_B=\sum_\kappa \varepsilon_\kappa({\cal N}_{1,\kappa}+{\cal N}_{2,\kappa}),
\end{equation}
where $[\overline{\Phi}_{i,\kappa},\Phi_{\kappa',j}]=0$ for
$\kappa\neq\kappa'$ and ${\cal N}_{i,\kappa}=\overline{\Phi}_{i,\kappa}\Phi_{i,\kappa}$.
For a given $\kappa$ the $SU_q(2)$ bosons are written 
in terms of boson operators $\phi_{i}$
and $\phi_{i}^{\dagger}$ with usual commutation relations:
$[\phi_i,\phi_j^{\dagger}]=\delta_{ij}$ as follows
\begin{eqnarray}
\Phi_j&=&(\phi_j^\dagger)^{-1} \{N_j\}q^{N_{j+1}},\\
\nonumber\\
\overline{\Phi}_j&=&\phi_j^\dagger q^{N_{j+1}}, \;\;\; j=1,2
\end{eqnarray}
leading to the  interacting boson Hamiltonian
\begin{eqnarray}
{\cal H}_B&=&\sum_{\kappa}\varepsilon_{\kappa}\{N_{1,\kappa}+N_{2,\kappa}\},\nonumber\\
&=&\sum_{\kappa}\frac{\varepsilon_{\kappa}}{q^2-1}\sum_{m=1}^{\infty}
\frac{2^m \ln^mq}{m!}\left(N_{1,\kappa}+N_{2,\kappa}\right)^m,
\end{eqnarray}
where $N_{i,\kappa}$ is the ordinary boson number operator and the bracket
$\{x\}=(1-q^{2x})/(1-q^2)$.  The grand partition function ${\cal Z}_B$ is given by
\begin{equation}
{\cal Z}_B=\prod_\kappa\sum_{n=0}^{\infty}\sum_{m=0}^{\infty}
e^{-\beta\varepsilon_\kappa\{n+m\}}e^{\beta\mu(n+m)},\label{Zb}
\end{equation}
which after rearrangement simplifies to the Equation
\begin{equation}
{\cal Z}_B=\prod_\kappa\sum_{m=0}^{\infty} (m+1)e^{-\beta\varepsilon_{\kappa}\{m\}}z^m,\label{Z}
\end{equation}
where $z=e^{\beta\mu}$ is the fugacity. 
\section{Low temperature behavior}\label{LT}

 In this Section we study the low temperature (high density) properties of the QGB
gas, which is represented by the partition function in Equation (\ref{Z}).  In particular, 
we will study the conditions under which BEC will occur.  In the thermodynamic
limit we write
\begin{equation}
\ln {\cal Z}_B=\ln\left(1+\sum_{n=1}^\infty(n+1)z^n\right)+\frac{4\pi V}{h^3}\int_0^{\infty}
 p^2\ln\left(1+\sum_{n=1}^{\infty}(n+1)e^{-\beta\{n\}\varepsilon}z^n\right) dp, \label{lnz}
\end{equation}
where, as in the ideal Bose case, the divergence of the ${\bf p}={\bf 0}$ term as $z\rightarrow 1$
has been taken into account by  splitting  $\ln {\cal Z}_B$ in a single term plus
an integration.  The series in the integrand of Equation (\ref{lnz}) diverges for $z=1$
and $q<1$, giving in this case no transition point.  For $q>1$, the series is more convergent
than the ideal Bose case, and BEC is expected to occur.  The average total number of particles
is given by
\begin{eqnarray}
\langle N\rangle &=& \frac{1}{\beta}\left(\frac{\partial\ln{\cal Z}_B}{\partial\mu}\right)_{T,V}\nonumber\\
&=&\langle N_0\rangle + \frac{4\pi V}{h^3}\left(\frac{2m}{\beta}\right)^{3/2}
\int_0^{\infty} x^2\frac{\sum_1^{\infty}(n+1)ne^{-\{n\}x^2}z^n}{f(z,q)} dx,
\end{eqnarray}
where the function $$f(z,q)=1+\sum_1^{\infty}(m+1)e^{-\{m\}x^2}z^m.$$ 
Expanding the integrand in powers of $z$ and integrating we obtain
\begin{equation}
\langle N\rangle = \langle N_0\rangle + \frac{V}{\lambda_T^3}\sum_{n=1}^{\infty} z^n S(n,q),\label{N}
\end{equation}
where $\lambda_T=\sqrt{h^2/2\pi mkT}$ and the coefficient $S(n,q)$
becomes, for $q=1$,  $S(n,1)=2/n^{3/2}$. Therefore, based on standard notation
we rewrite Equation (\ref{N}) as
\begin{equation}
 \langle N\rangle=\langle N_0\rangle+\frac{2V}{\lambda_T^3}g_{3/2}(z,q),\label{n}
\end{equation}
such that $g_{3/2}(1,1)=\zeta(3/2)$. In the $q\rightarrow 1$ limit, Equation (\ref{n}) corresponds to the case of
an ideal Bose gas with two species.  
  This model will exhibit Bose-Einstein condensation
for those values of the temperature or density such that
\begin{equation}
\lambda_T^3\geq 2\frac{V}{\langle N\rangle}g_{3/2}(1,q),
\end{equation}
where the critical temperature $T_c$ and critical density $\rho_c$ are given by the following 
equations 
\begin{equation}
T_c=\frac{h^2}{2\pi m k}\left(\frac{\langle N\rangle}
{2Vg_{3/2}(1,q)}\right)^{2/3}\label{temp},
\end{equation}
\begin{equation}
\rho_c=\frac{2g_{3/2}(1,q)}{\lambda_T^3}.
\end{equation}
Since for $q>1$ the function $g_{3/2}(1,q)<g_{3/2}(1,1)$, then the critical temperature $T_c$
for this model is larger than the critical temperature $T_c^{BE}$ for the ideal Bose gas.  For a given
density the two temperatures are related by
\begin{equation}
\frac{T_c}{T_c^{BE}}=\left(\frac{2.612}{g_{3/2}(1,q)}\right)^{2/3},
\end{equation}
giving then  $1\leq T_c/T_c^{BE}\leq 2.55$.  For the internal energy $U$ we have
\begin{eqnarray}
U&=&\frac{-\partial\ln {\cal Z}_B}{\partial\beta}+\mu \langle N\rangle \nonumber\\
&=& \frac{4V}{\sqrt{\pi}\lambda_T^3 \beta}\int_0^{\infty} x^4\frac{\sum_1^{\infty}(n+1)\{n\}e^{-\{n\}x^2}z^n}
{f(z,q)} dx.
\end{eqnarray}
Expanding the integrand in powers of $z$ leads to the equation
\begin{equation}
U=\frac{3}{2}\frac{2V}{\beta\lambda_T^3}g_{5/2}(z,q),
\end{equation}
with the identification $g_{5/2}(1,1)=\zeta(5/2)$.  

 Figure 1 shows a graph obtained 
from a numerical calculation 
of the functions $g_{3/2}(1,q)$ and $g_{5/2}(1,q)$ for $1\leq q\leq 6$.  
These functions decrease 
from its maximum value  $g_{3/2}(1,1)=2.612$ and $g_{5/2}(1,1)=1.341$ and remain
  practically constant for $q>3$.

\epsfxsize=400pt \epsfbox{g.epsf}

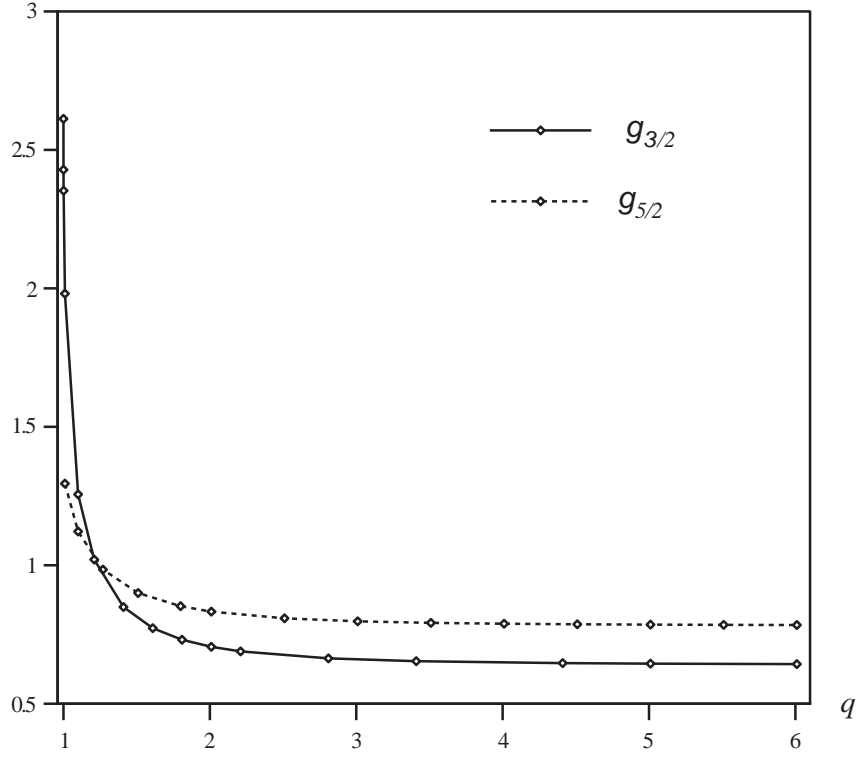
{\small  FIG. 1. The functions $g_{3/2}(1,q)$ and $g_{5/2}(1,q)$
, as defined in the text, for $1\leq q\leq 6$.  
The value of these functions decrease from its maximum value at $q=1$,
 and  remains approximately  constant for $q>3$}\\
\\
The heat capacity
is  written in terms of the internal energy according to
the general expression
\begin{eqnarray}
C_v&=&\left(\frac{\partial U}{\partial T}\right)_V\nonumber\\
&=&\frac{5}{2}\frac{U}{T}+ \left(\beta\mu'-\frac{\mu}{kT^2}\right)z\frac{\partial}{\partial z}U,
\end{eqnarray}
where $\mu'=\partial\mu/\partial T$.  For  values of the temperature
very
close to $T_c$ the chemical potential $\mu(T_c^-)=0$ and
its derivative $\mu'(T_c^-)=0$, and therefore if $\mu'(T_c^+)\neq 0$
there will be
a discontinuity in the heat capacity according to
\begin{eqnarray}
\Delta C_v&=&C_v(T_c^-)-C_v(T_c^+)\nonumber\\
&=&-\beta\mu'(T_c^+)\left(z\frac{\partial U}{\partial z}\right)_{z=1}.
\end{eqnarray}
A simple way \cite{BPK} to find $\mu'(T_c^+)$ is by first
considering that for a closed system $\partial \langle N\rangle/\partial T=0$,
such that for $T>T_c$, $\langle N_0\rangle=0$ so from Equation (\ref{n}),
we obtain
\begin{equation}
\mu'(T_c^+)=-\frac{3}{2}k\left(g_{3/2}(z,q)/
z\frac{\partial g_{3/2}(z,q)}
{\partial z}\right)_{z=1}.\label{mu'}
\end{equation}
At $q=1$ the denominator  of Equation (\ref{mu'}) becomes the divergent series
$\sum_1^\infty k^{-1/2}$, and then $\Delta C_v=0$.
 The denominator converges for
all values of $q>1$, showing  that the
heat capacity exhibits a gap at the critical temperature.

In particular, for low temperatures such that
$T<T_c$ we have that the chemical potential $\mu(T<T_c)=0$ and its
derivative $\frac{\partial\mu}{\partial T}(T<T_c)=0$, giving for the heat capacity
\begin{equation}
C_v=\frac{15}{2}\frac{kV}{\lambda_T^3}g_{5/2}(1,q),
\end{equation}
such that after inserting Equation (\ref{temp}) the heat capacity becomes
\begin{equation}
C_v=\frac{15}{4}k \langle N\rangle \frac{g_{5/2}(1,q)}{g_{3/2}(1,q)}\left(\frac{T}{T_c}\right)^{3/2}\label{C1}.
\end{equation}
The dependence of the function $g_{5/2}/g_{3/2}$ on $q$ will indicate how much
the behavior of the heat capacity for this model departs from the ideal
Bose case.

In order to find the heat capacity for temperatures $T>T_c$,
 we expand the $\ln {\cal Z}_B$ in powers of
the fugacity $z$.  After performing the elementary integrations,
the first few terms read
\begin{equation}
\ln {\cal Z}_B=\frac{V}{\lambda_T^3}\left(2z+4\delta(q)z^2+...\right),
\end{equation}
where $\delta(q)=\frac{1}{4}\left([3/(1+q^2)^{3/2}]-(1/\sqrt{2})\right)$ .
Calculating the average number of particles and reverting the equation,
we find,  the fugacity and the internal energy as functions of $\langle N\rangle$
\begin{equation}
z=\frac{\langle N\rangle\lambda_T^3}{2V}\left(1-\frac{2\lambda_T^3\delta(q)
\langle N\rangle}{V}+...\right),
\end{equation}
\begin{equation}
U=\frac{3\langle N\rangle}{2\beta}\left(1-\frac{\lambda_T^3 \delta(q)\langle N\rangle}{V}+
...\right).
\end{equation}
\epsfxsize=400pt \epsfbox{c.epsf}

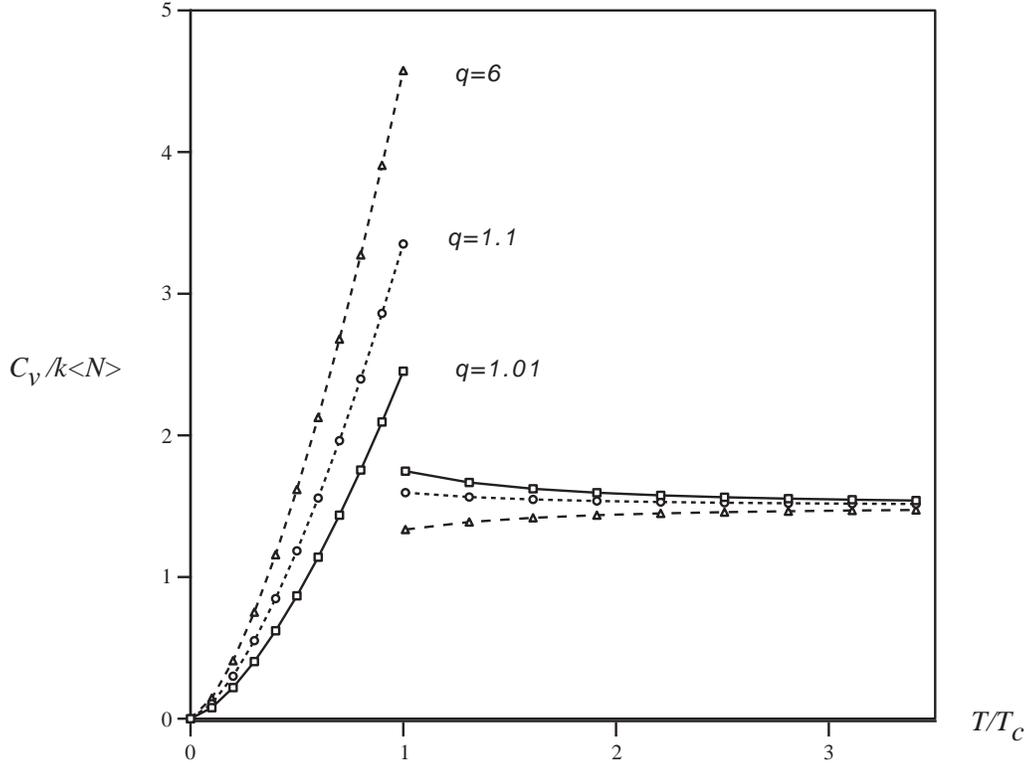
{\small  FIG. 2. The heat capacity $C_v$ as a function of the temperature
for several values of the 
parameter $q$.  The
gap at the condensation temperature increases with  $q$ up to a value
$\Delta C_v=3.24 k \langle N\rangle$ at $q=6$ and remains constant
for $q>6$.}
\\

The heat capacity is then given by
\begin{equation}
C_v=\frac{3k \langle N\rangle}{2}\left(1+\frac{\lambda_T^3 \delta(q)\langle N\rangle}
{2V}+...\right),
\end{equation}
such that with the use of Equation (\ref{temp}) it becomes
\begin{equation}
C_v=\frac{3k \langle N\rangle}{2}\left(1+\delta(q) g_{3/2}(1,q)\label{C2}
\left(\frac{T_c}{T}\right)^{3/2}+...\right).
\end{equation}
According to Equations (\ref{C1}) and (\ref{C2}) the gap in the heat capacity
  is then given by the Equation
\begin{equation}
\Delta C_v\mid_{T=T_c}\approx k\langle N\rangle\left(\frac{15 g_{5/2}(1,q)}
{4 g_{3/2}(1,q)}-\frac{3}{2}\left(1+\delta(q) g_{3/2}(1,q)\right)\right).
\end{equation}

Figure $2$ shows the discontinuity of the heat capacity $C_v$ for
different values of $q$, and shows that the gap  is more
sensitive to small deviations of $q$ from the ideal Bose
case $q=1$. The gap increases with the value of the parameter $q$ and remains approximately constant
for $q>3$. Thus, in this model, the onset of BEC becomes
a second order phase transition.

At  low temperatures $T<T_c$, the chemical potential $\mu=0$, giving for the entropy  
and the equation of state
\begin{eqnarray}
S &=&\frac{U}{T}+k \ln{\cal Z}_B\nonumber\\
&=&\frac{5kV}{\lambda_T^3}g_{5/2}(1,q),\\
 \frac{p}{kT}&=&\frac{2}{\lambda_T^3} g_{5/2}(1,q).
\end{eqnarray}
According to Figure 1, the entropy has the interesting property that
for
a given temperature $T<T_c$ it  acquires its maximum value at $q=1$.  
This same feature was also found at
low temperatures for the case of a quantum group fermion gas \cite{U1}.
Then, the interactions introduced by the $SU_q(2)$ symmetry are such that
they decrease the entropy below the one corresponding to the standard case $q=1$.
\section{Conclusions}

In this paper we studied the low temperature behavior of a quantum group gas
with $SU_q(2)$ symmetry in the thermodynamic limit.  Our results indicate, that the boson interactions
introduced by the quantum symmetry are such that this system exhibits Bose-Einstein
condensation for $q>1$. 

The  interaction terms  are fixed by the quantum symmetry of
the Hamiltonian, which are the values $N$ and $q$ of $SU_q(N)$.  For purposes
of simplicity, we set $N=2$ and analyzed the effect of varying  $q$. 
The parameter $q$, which
in the boson representation controls the strength of the interactions,
 plays the role of increasing the condensation
temperature, reducing the entropy, and producing a gap in the heat capacity at the onset of the BEC.  
Our results are written in terms of the  functions $g_{3/2}(1,q)$ and 
$g_{5/2}(1,q)$  introduced in the text.  These functions decrease rapidly from their respective values  
 $\zeta(3/2)$ and $\zeta(5/2)$ at $q=1$, and become approximately constant for $q>3$.
Therefore, the properties of this model will be more sensitive to those $q$ values
which are small deviations from the ideal Bose case $q=1$. A simple numerical check
shows that the critical temperature for $He^4$, $T_c\approx 2.20 K$, corresponds
to $q=1.02$.  Even more interesting is the fact that the  same value $q=1.02$
fits very well the gap in the heat capacity of a
dilute gas of $^{87}Rb$ atoms, as reported in  Reference \cite{E}, although
a realistic comparison with experimental results requires  to consider, in addition, an
external harmonic potential and the corrections due to a small number
$(\approx 20,000)$ of particles. Another
interesting feature of this model is that at high temperatures and $q>1.78$
it behaves as an interacting fermion gas \cite{U2} and, on the other hand,
at low temperatures and $q>1$ it exhibits BEC.

The system studied in this paper is described by the simplest quantum group
invariant Hamiltonian, and the calculations were made in the thermodynamic limit and with no external potential.
   Some recent studies for the case of an ideal Bose gas inside a  trapping potential, 
in the thermodynamic limit \cite{BPK,BK,HR}; and
with a finite number of particles \cite{GH,Mu,HHR} 
show the effect of these corrections on the condensation temperature and
heat capacity. Therefore, since in our case, the interactions are fixed
by the  $SU_q(2)$ symmetry of the Hamiltonian, a natural continuation
of our work will be to include those corrections such that a more direct
comparison to  experimental results can be made.

\end{document}